\documentclass[11pt,twoside,onecolumn]{article}
\usepackage[]{latexsym}
\usepackage{epsfig}
\usepackage{amsmath,amssymb}
\usepackage{graphicx}
\usepackage{makeidx}
\newcommand{\be}{\begin{eqnarray}}
\newcommand{\ee}{\end{eqnarray}}

\title{\bf Quantum Black Hole Decay with MadGraph}

\author{G.L.~Alberghi$^{a}$\thanks{alberghi@bo.infn.it}
\\
\null
\\
$^a${\em Istituto Nazionale di Fisica Nucleare, Sezione di Bologna}
\\
{\em viale B.~Pichat~6/2, 40127 Bologna, Italy}
\\
}
\begin{document}
\maketitle
\begin{abstract}
We begin the investigation of the decay of quantum black holes with masses of the order of the TeV,
possibly produced at the Large Hadron Collider, within the framework of MadGraph 5,
a standard particle physics Monte Carlo event generator.
We write an effective Lagrangian to describe the black hole decay and examine both the possibility of a complete decay 
into standard model particles with or without the associated production of a very massive stable neutral remnant.

% As a first step toward the impementation of Black Hole decay in particle physics Monte Carlo generators,
% we begin the study of Black Hole decay by employing a standard Monte Carlo generator, 
% namely MadGraph5.  

%We investigate possible signatures of long-lived (or stable) charged black holes
%at the LHC.
%In particular, we find that both charged and neutral remnants could in some cases
%be very clearly distinguished from the background events because of their low speed. 
%For the charged remnants, we then provide an estimate of  the energy released
%inside a typical detector using the Bethe-Bloch formula.
%
\end{abstract}
\setcounter{page}{1}
\section{Introduction}
\setcounter{equation}{0}
It is now well appreciated that the scale at which quantum gravity effects
become comparable in strength to the forces of the Standard Model (SM)
of particle physics could be well below the traditional Planck mass of $10^{19}\,$GeV,
and potentially in the TeV~range~\cite{arkani,RS,Calmet:2008tn}.
In fact, models with low scale quantum gravity allow for a fundamental scale
of gravity as low as the electroweak scale, $M_{\rm G}\simeq 1\,$TeV.
In this case microscopic black holes might be produced particle in accelerators (see, e.g., Ref.~\cite{cavaglia}).
Until recently, only semi-classical black holes, which  decay via the
Hawking radiation~\cite{hawking}, had been considered.
These black holes, whose standard description is based on the canonical Planckian
distribution for the emitted particles, have a very short life-time
of the order of $10^{-26}\,$s~\cite{dimopoulos}.
The creation of semi-classical black holes in collisions of high energetic particles
is well understood~\cite{D'Eath:1992hb,Eardley:2002re,Hsu:2002bd}.
Our understanding of this phenomenon thus goes way beyond the naive
hoop conjecture~\cite{hoop} used in the first papers on the topic and
the Large Hadron Collider was able to set some bounds on the Planck mass,
searching for semi-classical black holes~\cite{CMS,park,Aad:2014gka} by using the predictions derived by
using several simulation tools, available to describe black hole
production and evaporation at colliders~\cite{harris,catfish,cha2,blackmax,charybdis2},
%Several simulation tools are also available to describe semi-classical BH
%production and evaporation at colliders~\cite{harris,catfish,cha2,blackmax,charybdis2},
\par
Recently, it has been pointed out that, besides semi-classical blackholes,
which appear to be difficult to produce at colliders, as they might require energies 5 to 20
times larger than the Planck scale, quantum black holes, could be instead copiously
produced~\cite{Calmet:2008dg,Calmet:2011ta,Calmet:2012cn,Calmet:2012mf}.
These black holes are non-thermal objects with masses close to the Planck scale,
and might resemble strong gravitational rescattering events~\cite{Meade:2007sz}. 
In Ref.~\cite{Calmet:2008dg}, non-thermal quantum black holes were assumed
to decay into only a couple of particles.
However, depending on the details of quantum gravity, the smallest quantum black holes
might be stable and would not decay at all.
The existence of remnants, i.e.~the smallest stable black holes, have been considered previously
in the literature~\cite{Koch:2005ks,Hossenfelder:2005bd}.
And, most recently, the production of neutral and integer charged semi-classical
remnant black holes have been simulated in Ref.~\cite{Alberghi:2013hca}.
\par
By assuming a decay into a maximum of four standard model particles, and allowing also the possibility
of producing a stable massive neutral remnant, we were able to write down an effective Lagrangian
for the interaction between the quantum black hole and the standard model particles and the remnent.
This allows a direct implementation of the model in MadGraph 5 (\cite{Alwall:2011uj}) and thus the use of
a its powerful machinery. % standard phenomenological analysis.
In particular we are able to study the properties of quantum black hole decay events,
considering bot the possibility that its decay products contain or not a massive neutral remnant.
%This article is closely related to in correlation and following the work of \cite{Alberghi:2013hca}, 
%where we studied the particle speed distributions in search for a possible signature
%to differentiate charged remnants from Standard Model particles.

%**** In fact, we will show that remnant black holes could be produced with relativistic
%factors much smaller than those of Standard Model particles. **** 
%For the charged remnant BHs, we also estimate the typical energy loss inside
%a detector using the Bethe-Bloch equation and the distributions in speed.
\par
\par
The paper is organised as follows: 
in the Section~\ref{production}, we briefly review black hole production in the semi-classical regime 
and extrapolate possible behaviours in the quantum regime;
in Section~\ref{model} we describe the model impelemented in MadGraph 5; 
%mostly arguing on the possible existence of BHs with electric charge
%(see also Appendix~\ref{BWBH})
in Section~\ref{results} we complete our phenomenological analysis; 
we finally comment, conclude and argue about further developments in Section~\ref{conclusions}.

%we summarise the main results for remnant BHs
%from Ref.~\cite{Bellagamba:2012wz}, and complete our analysis with the
%study of their distributions in speed.
%We confront these distributions with the analogous distributions
%for Standard Model particles produced in the same events,
%and in events with $t\,\bar t$ but no BHs, which represent the background for processes
%of interest here.
%We also estimate the typical energy released in a detector by charged remnants;
%
%
%
\section{Black hole production}
\label{production}
\setcounter{equation}{0}
In this work, we start from the possibility that remnant black holes could be the end-point of the
Hawking evaporation, but they might also be produced directly without going
through the usual evaporation process but through a decay process.
The physics of the latter is very similar to that described in Ref.~\cite{Calmet:2008dg},
with the important exception that they would be stable.
\par
In a proton-proton collider such as the LHC, black holes would be produced
by quarks, anti-quarks and gluons and would thus typically carry both a QED
and a SU(3)$_c$ charge, namely
\begin{itemize}
\item[a)] ${\bf 3} \times {\bf \overline 3}= {\bf 8} + {\bf 1}$
\item[b)] ${\bf 3} \times {\bf 3}= {\bf 6} + {\bf \overline 3}$
\item[c)] ${\bf 3} \times {\bf 8}= {\bf 3} + {\bf \overline 6}+ {\bf 15}$
\item[d)] ${\bf 8} \times {\bf 8}= {\bf 1}_S + {\bf 8}_S+ {\bf 8}_A+{\bf 10} + {\bf \overline{10}}_A+ {\bf 27}_S$
\end{itemize}
Most of the time, black holes will thus be created with a SU(3)$_c$ charge and
come in different representations of SU(3)$_c$, as well as QED charges.
It is also likely that their masses are quantized, as described in Ref.~\cite{Calmet:2012cn}.
Remnant quantum black holes can therefore be classified according to representations of SU(3)$_c$.
Let us further note that in Refs.~\cite{Calmet:2008dg,Calmet:2011ta,Calmet:2012cn,Calmet:2012mf}
we did not expect that non-thermal quantum black holes would ``hadronize" before decaying
(since the QCD length scale is 200$^{-1}\,$MeV, whereas that of quantum gravity
in these scenario is at most 1000$^{-1}\,$GeV).
However, since the black holes we consider here are stable, we expect that they will hadronize,
i.e.~absorbe a particle charged under $SU(3)_c$ after traveling over a distance of some
$200^{-1}\,$MeV and become an $SU(3)_c$ singlet.
The hadronization process could lead to remnants with zero electric charge,
however, they could still have a (fractional) QED charge.
% and the phenomenology
%is possibly rather different from the one envisaged in Ref.~\cite{Bellagamba:2012wz}.
%Quantum BHs can thus have the following QED charges:
%$\pm 2/3, \pm 1/3, \pm, \pm 4/3$ and $\pm 1$.
%If the BH is fast moving, it is likely to hadronize in the detector, whereas if it is slowly moving,
%this is likely to happen before it reaches the detector.
%The hadronization process, however, does not guarantee that the charge of the resulting
%small BH will be an integer.
%Indeed, in most cases, it will still be a fractional charge, depending on how many colored
%particles are absorbed to neutralize the BH from the color point of view.
\par
The production cross section is in any case extrapolated from the semi-classical
regime and assumed to be accurately described by the geometrical cross section formula.
The horizon radius, which depends on the number $d$ of extra-dimensions, is given by
\be
R_{\rm H}
=\frac{\ell_{\rm G}}{\sqrt{\pi}}\,
\left(\frac{M}{M_{\rm G}}\right)^{\frac{1}{d+1}}
\left(\frac{8\,\Gamma\left(\frac{d+3}{2}\right)}{d+2}
\right)^{\frac{1}{d+1}}
\ ,
\ee
where $\ell_{\rm G}=\hbar/M_{\rm G}$ is the fundamental gravitational
length associated with $M_{\rm G}$,
$M$ is the black hole mass, $\Gamma$ the Gamma function.
At the LHC, a black hole could form in the collision of two partons, i.e.~the quarks, anti-quarks
and gluons of the colliding protons.
The total black hole cross section, 
\be
\left.\frac{d\sigma}{d M}\right|_{pp\to BH+X}
=\frac{dL}{d M}\,\sigma_{\rm BH}(ab\to BH; \hat s=M^2)
\ ,
\ee
can be estimated from the geometrical hoop conjecture~\cite{hoop},
so that 
\be
\sigma_{\rm BH}(M)\approx \pi\,R_{\rm H}^2
\ ,
\ee
and
\be
\frac{dL}{d M}=\frac{2\,M}{s}\,\sum_{a,b}\int_{M^2/s}^1
\frac{dx_a}{x_a}\,f_a(x_a)\,f_b\left(\frac{M^2}{s\,x_a}\right)
\ ,
\ee
where $a$ and $b$ represent the partons which form the black hole,
$\sqrt{\hat s}$ is their centre-mass energy and $f_i(x_i)$ are
parton distribution functions (PDF)
and $\sqrt{s}$ the LHC centre-mass collision energy
(up to $8\,$TeV presently, with a planned maximum of $14\,$TeV).
\par
%As an example, in the following we shall consider $\sqrt{s}=14\,$TeV,
%and $\sigma_{\rm BH}\simeq 39.9\,$fb for $M_{\rm G}=3.5\,$TeV in $D=6$
%dimensions~\cite{Bellagamba:2012wz}.
%In the first few months of the future LHC run, a luminosity $L\simeq 10\,$fb$^{-1}$
%should be reached, and one can therefore expect a total of about $390$ black hole events.
%
%
%
\section{The effective model}
\label{model}
\setcounter{equation}{0} 
MadGraph 5 allows the implementation of new models of particle physics by simply writing down a Lagrangian.
A dedicated routine computes the deriving Feynman diagrams, allowing the calculation of matrix elements,
and consequently the decay width.
This first step in the analysis of quantum black hole decay is the analysis we describe in this section.
In our model, the quantum black hole is implemented as a heavy scalar particle coupling with 
universal coefficents to standard model particles.
In particular, as a first step in this new line of research, we allowed the possibility for the black hole
to decay into a maximum of four Standard Model Particles, with or without the production of a stable
neutral remnant. We chose the mass of the remnant to be 500 GeV lower that that of the decaying black hole,  
in order to explore the possibility of having an high mass neutral stable particle in the final state,
maximizing the importance of the new decay channel.
In particular we examined the case for quantum blck holes of mass 1.5, 2, 2.5, 3, 3.5 TeV.
The interaction terms are of the form,
\begin{figure}
  \includegraphics[width=0.5\textwidth]{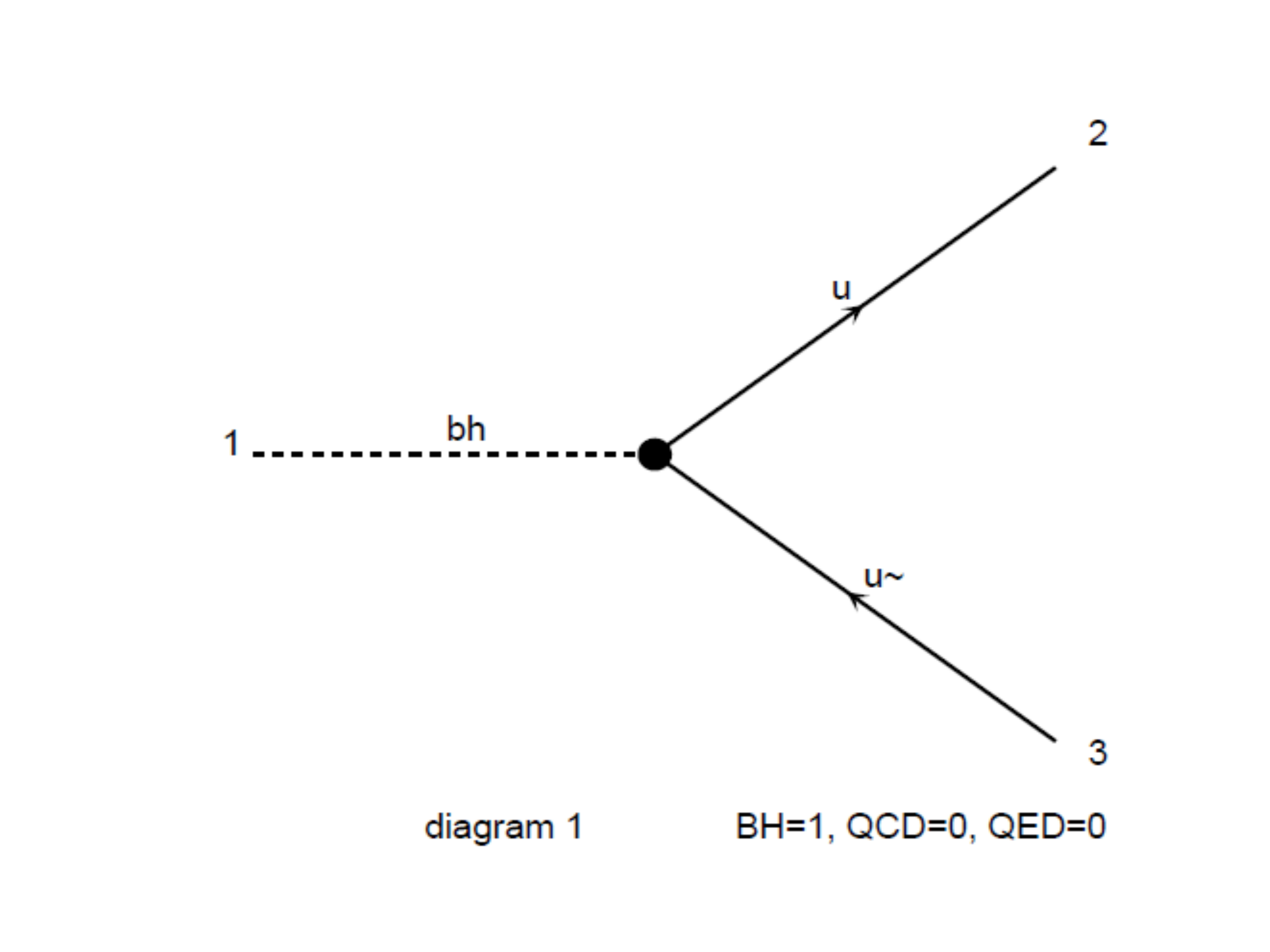}
  \includegraphics[width=0.5\textwidth]{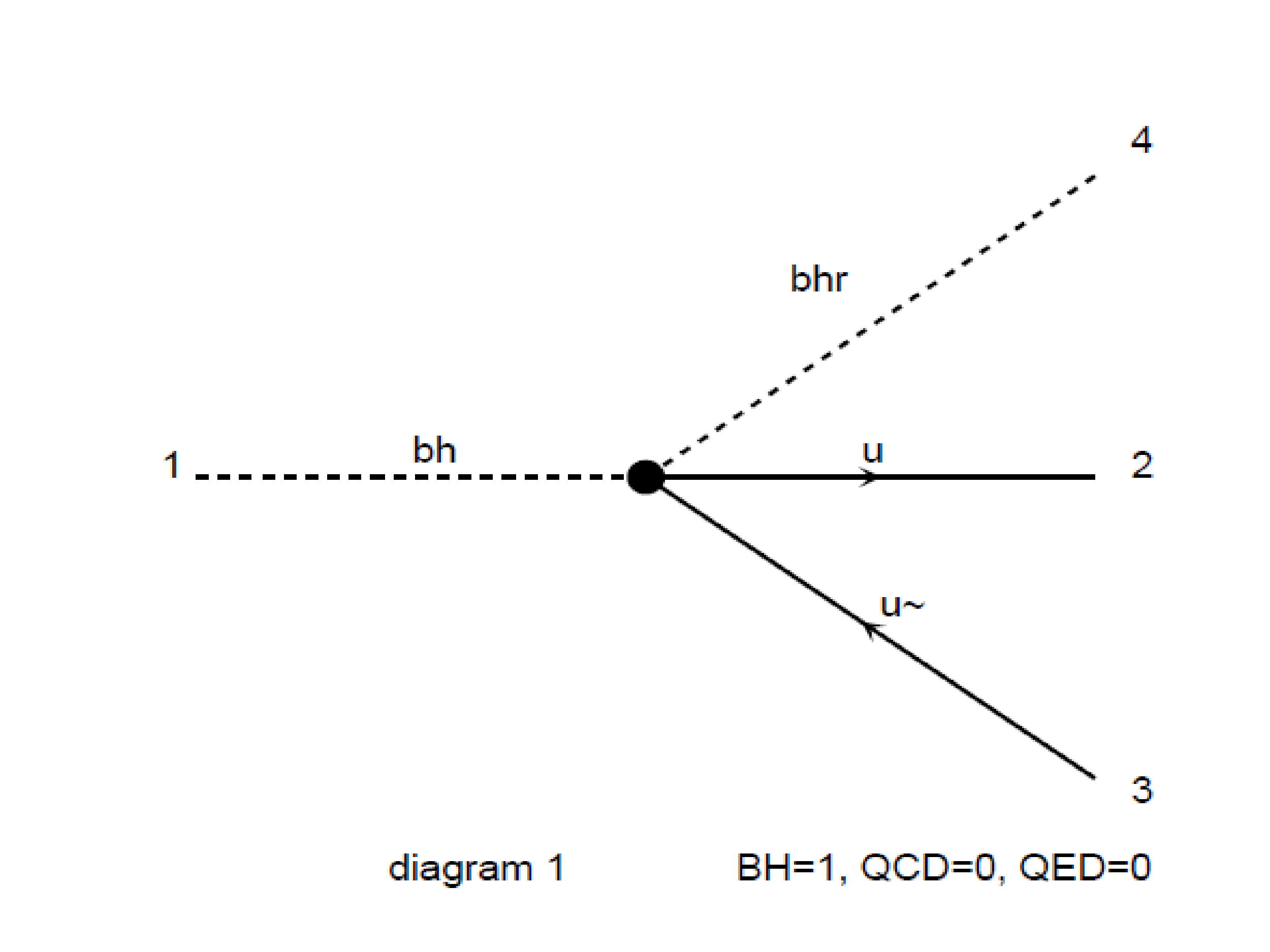}   
  \caption{The Feynman diagrams corresponding to the interaction term of Eq.\ref{interaction}}
  \label{fig:interaction}
\end{figure}
%\begin{figure}
%\center
%\includegraphics[width=0.6\textwidth]{35PTbhr_all.png}
%\includegraphics[width=0.6\textwidth]{35PTbhr_bhr.png}
%\includegraphics[width=0.6\textwidth]{35PTbhr_nobhr.png}
%\caption{PTbhr}
%\end{figure}
%\begin{figure}
%\centerline{\begin{tabular}{ r}
%\centerline{\hbox{
% \includegraphics[width=0.7\textwidth]{bh_uu_tagliato.pdf}
% \includegraphics[width=0.7\textwidth]{bh_uubhr_tagliato.pdf}
% }
%\end{tabular}}
%\caption{Feynman diagrams corresponding to Eq.\cite{interaction}}
%\label{fig:feynman}}
%\end{figure} 
(we take as an example the coupling with two quarks, which correspond to the diagrams of Fig.\ref{fig:interaction}, 
as representd by MadGraph)
\be
 L_{int} \supseteq
{1 \over  2} \, g^{(3)}_{BH} \,   \bar{u} \,  u  \, \text{bh} + {1 \over  2} \, g_{BH}^{(4)}  \, \bar{u} \, u \, \text{bh  bhr} 
\label{interaction}
\ee
where bh represents the black hole, bhr is the remnant, $g^{(3)}_{BH}$ 
is the universal coupling constant of the black hole with two SM particles and the 
corresponding $g_{BH}^{(4)}$ is the term allowing for the production of the remnant.
As already said, the Lagrangian allows the possibility of producing up to four SM particles,
thus four universal coupling constants were introduced and all possible terms were added 
to the interaction Lagrangian.
The value of the coupling constants were chosen in order to give rise to decay width comparable to the so called democraticity 
of the black hole decay, and had to be slightly adjusted depending on the mass of the decaying black hole,
in order for all the decay channels to have comparable width.
Clearly, in the following steps of the analysis, such a dependance shuld be investigated in more detail.
In table \ref{tab:parameters} we summarize the values of the interacion constants and the corresponding decay
width for the model we are considering.
\begin{table}[h]
\renewcommand{\arraystretch}{1.5}
\centering
\begin{tabular}{|l|l|l|l|l|l|l|}
  \hline
  BH Mass (TeV) & Remnant Mass (Tev) & $g_{BH}^{3}$ & $g_{BH}^{4} $ &  $g_{BH}^{5} $ & $g_{BH}^{6} $ & Decay Width (GeV) \\ \hline 
 1.5 &  1.0 & $10 ^{-2}$ & $10 ^{-2}$ & $3 \cdot 10 ^{-6}$ & $ 3 \cdot 10 ^{-5}$  &  167  \\ \hline
 2.0 &  1.5 & $10 ^{-2}$ & $10 ^{-2}$ & $  10 ^{-6}$ & $ 3 \cdot 10 ^{-5}$  &  124  \\ \hline
 2.5 &  2.0 & $10 ^{-2}$ & $10 ^{-2}$ & $  10 ^{-7}$ & $ 5 \cdot 10 ^{-5}$  &  89  \\ \hline
 3.0 &  2.5 & $10 ^{-2}$ & $10 ^{-2}$ & $  10 ^{-7}$ & $ 5 \cdot 10 ^{-5}$  &  74  \\ \hline
 3.5 &  3.0 & $10 ^{-2}$ & $10 ^{-2}$ & $  10 ^{-7}$ & $ 5 \cdot 10 ^{-5}$  &  87  \\ \hline
\end{tabular}
\caption{The model parameters}
\label{tab:parameters}
\end{table}

%%\centerline{\hbox{
%\epsfig{figure=betaremn-fullfalse-14tev_3.5_6.eps,height=6cm,clip=}
%\epsfig{figure=remnmass-fullfalse-14Tev_3.5_6.eps,height=6cm,clip=}
%}}
%\end{tabular}}

\section{The Decay}
\label{results}
\begin{figure}
\center
\includegraphics[width=0.6\textwidth]
{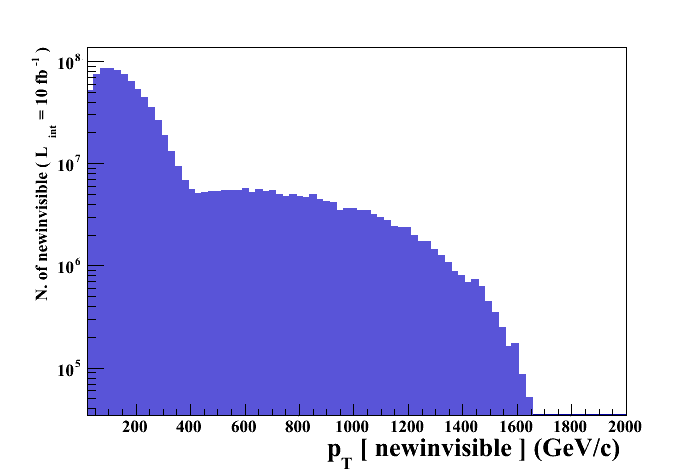}
\includegraphics[width=0.6\textwidth]
{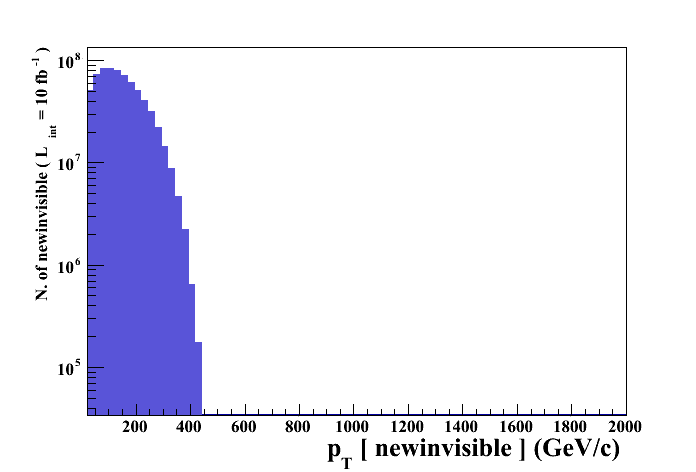}
\includegraphics[width=0.6\textwidth]
{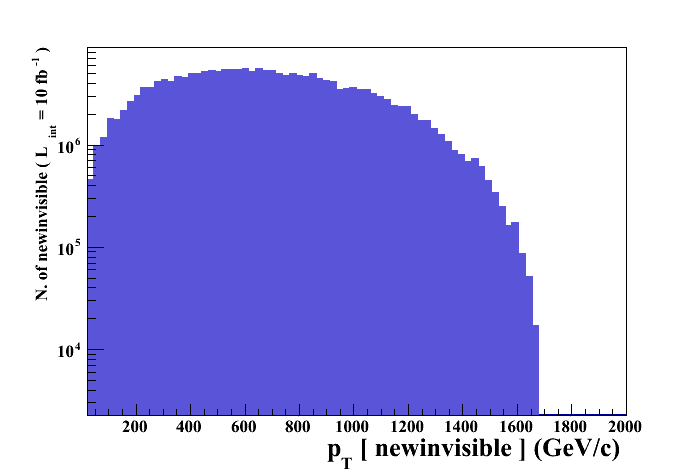}
\caption{Missing transverse momentum for 3.5 TeV black hole decay events for all events (upper), for events with a remnant (middle) and for events without remnant (lower)}
\label{Pt35}
\end{figure}

\begin{figure}
\center
\includegraphics[width=0.6\textwidth]
{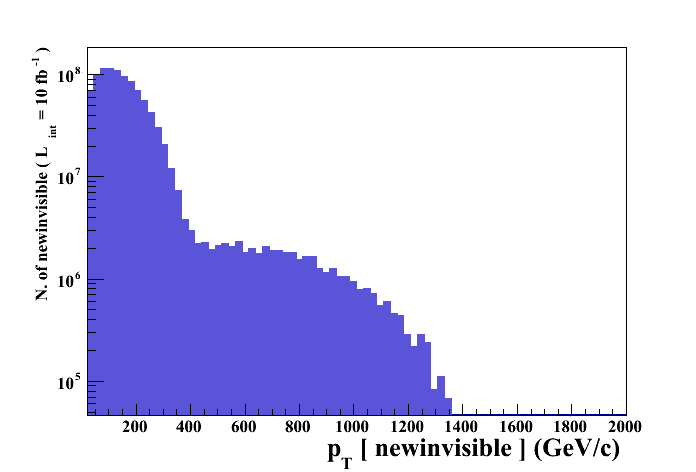}
\includegraphics[width=0.6\textwidth]
{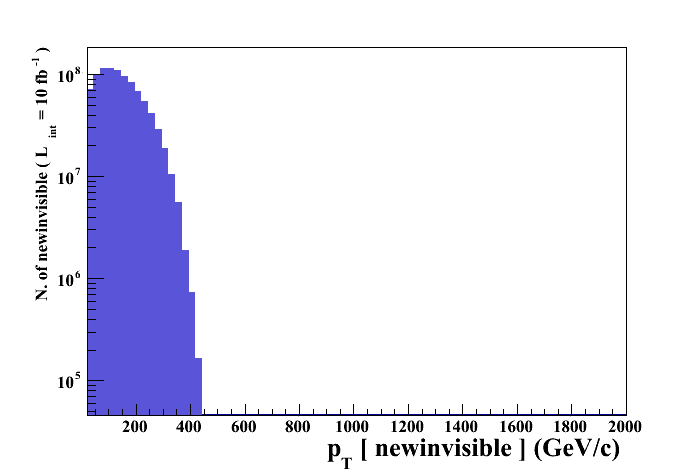}
\includegraphics[width=0.6\textwidth]
{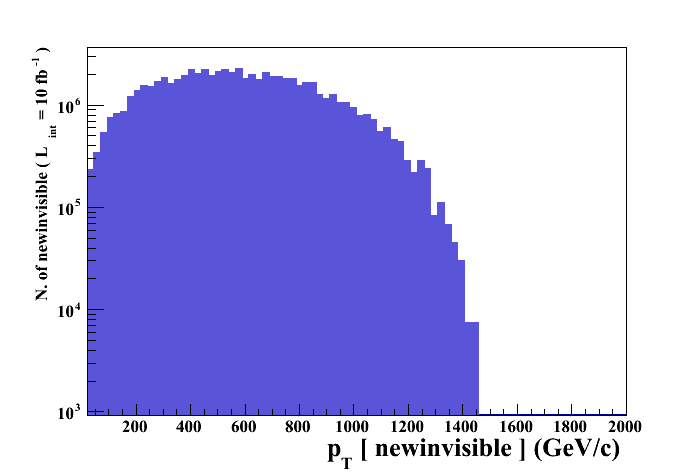}
\caption{Missing transverse momentum for 3 TeV black hole decay events for all events (upper), for events with a remnant (middle) and for events without remnant (lower)}
\label{Pt30}
\end{figure}

\begin{figure}
\center
\includegraphics[width=0.6\textwidth]
{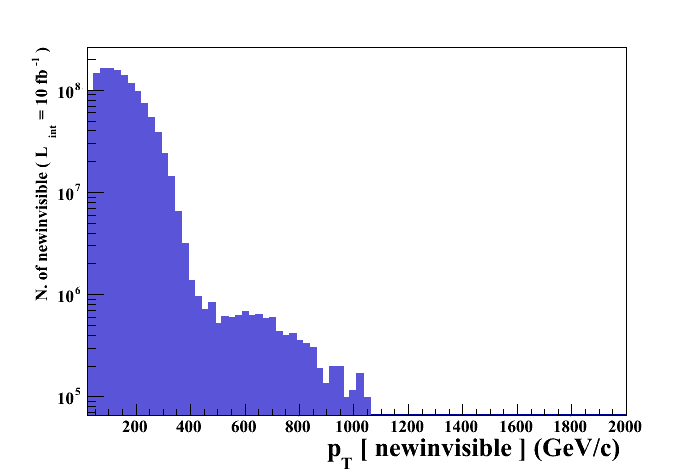}
\includegraphics[width=0.6\textwidth]
{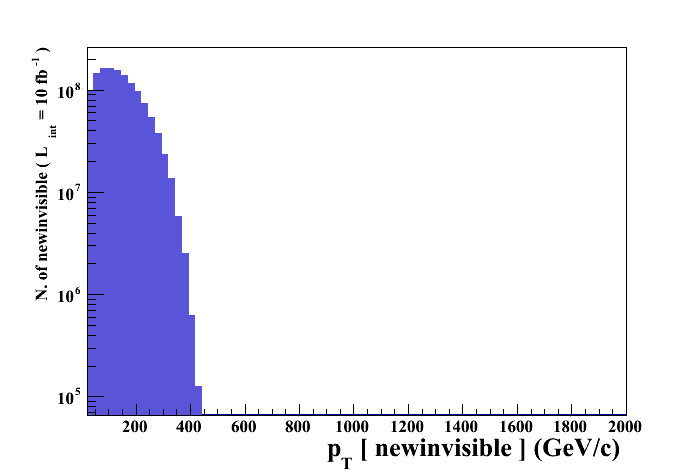}
\includegraphics[width=0.6\textwidth]
{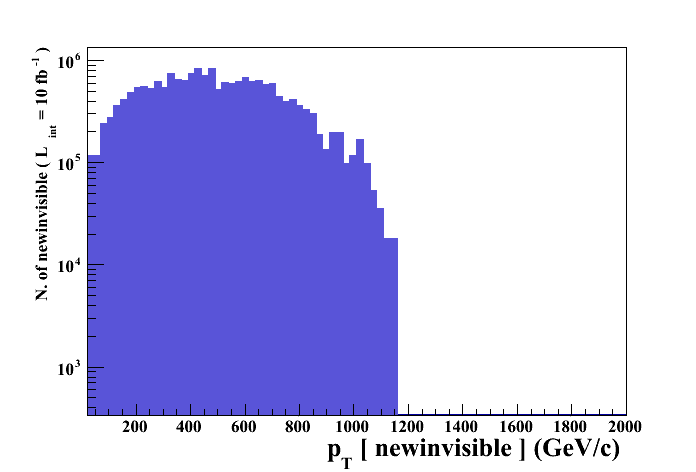}
\caption{Missing transverse momentum for 2.5 TeV black hole decay events for all events (upper), for events with a remnant (middle) and for events without remnant (lower)}
\label{Pt25}
\end{figure}

\begin{figure}
\center
\includegraphics[width=0.6\textwidth]
{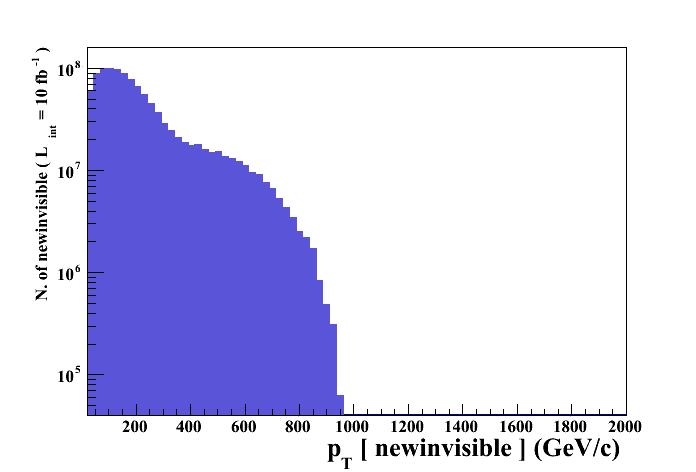}
\includegraphics[width=0.6\textwidth]
{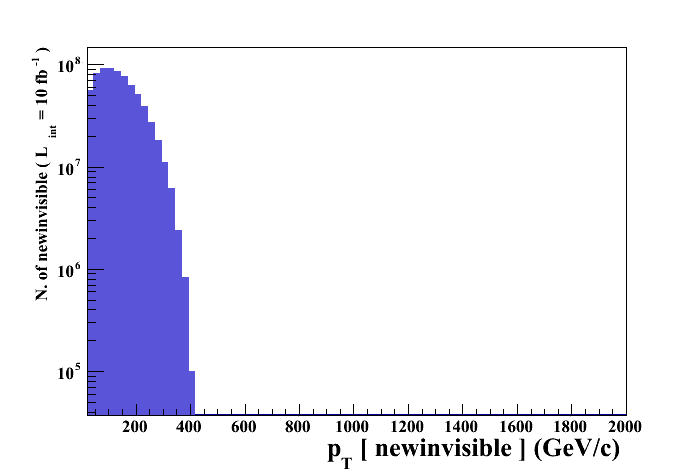}
\includegraphics[width=0.6\textwidth]
{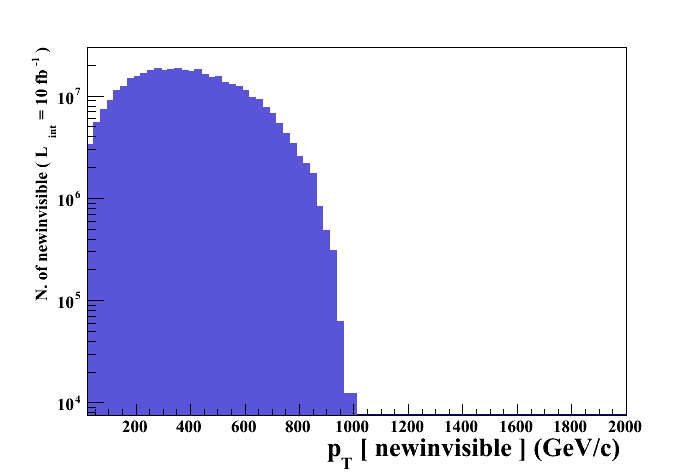}
\caption{Missing transverse momentum for 2 TeV black hole decay events for all events (upper), for events with a remnant (middle) and for events without remnant (lower)}
\label{Pt20}
\end{figure}

\begin{figure}
\center
\includegraphics[width=0.6\textwidth]
{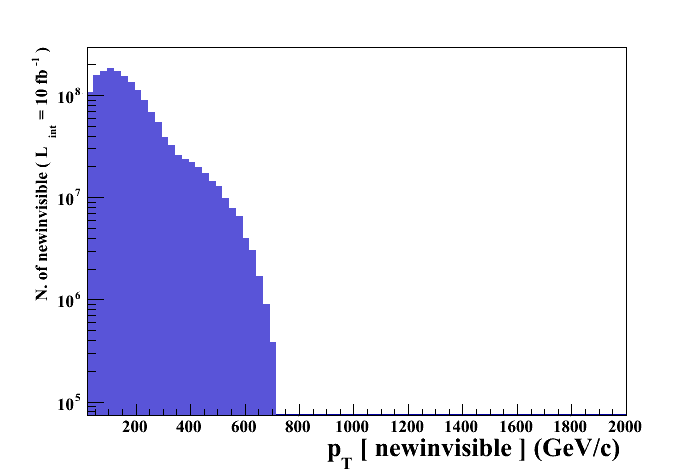}
\includegraphics[width=0.6\textwidth]
{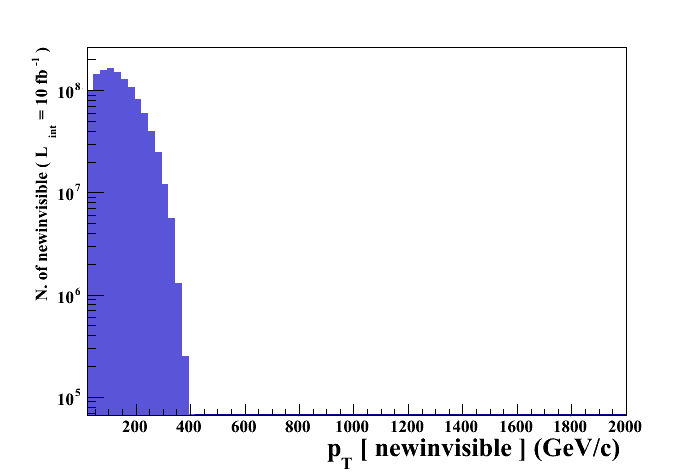}
\includegraphics[width=0.6\textwidth]
{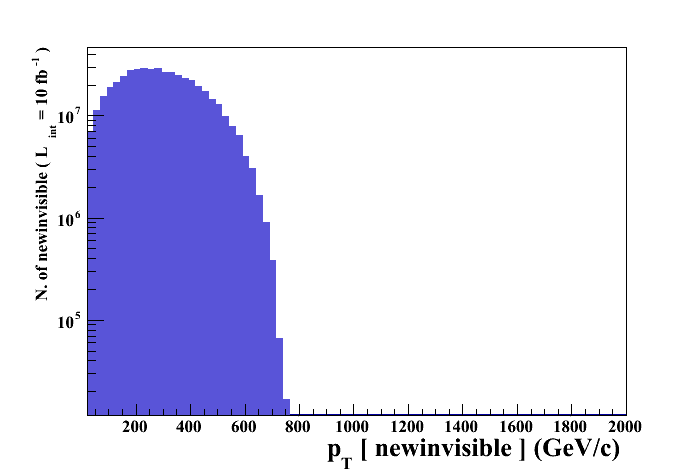}
\caption{Missing transverse momentum for 1.5 TeV black hole decay events for all events (upper), for events with a remnant (middle) and for events without remnant (lower)}
\label{Pt15}
\end{figure}

\begin{figure}
\center
\includegraphics[width=0.6\textwidth]
{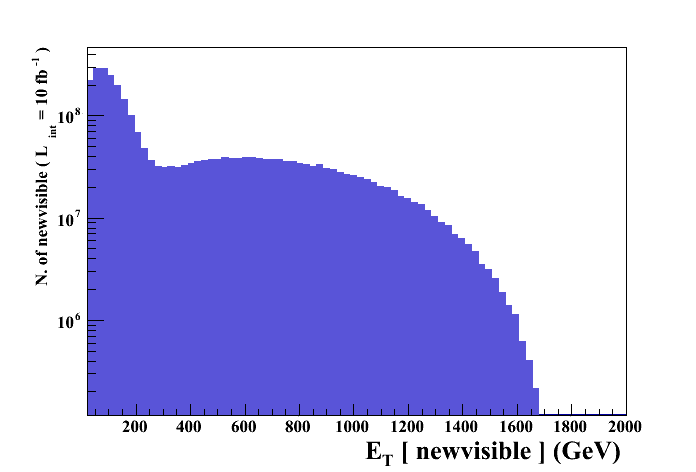}
\includegraphics[width=0.6\textwidth]
{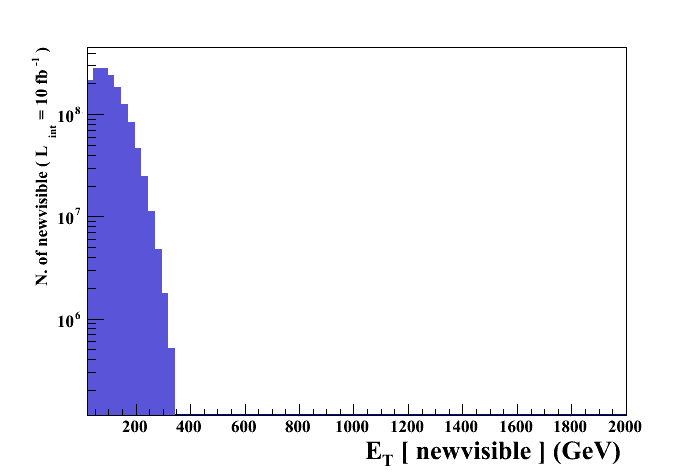}
\includegraphics[width=0.6\textwidth]
{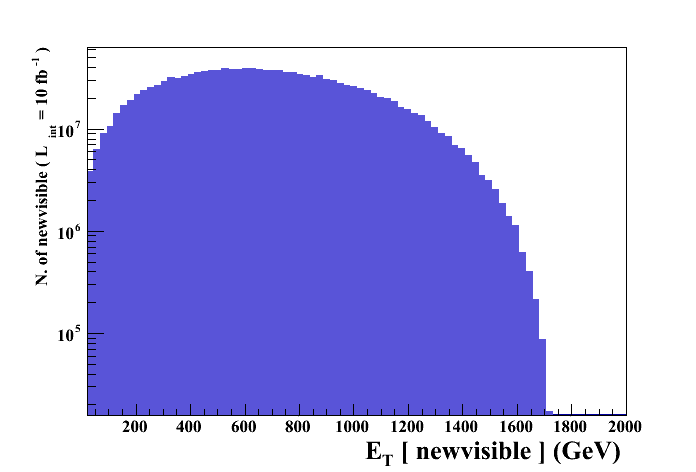}
\caption{Missing transverse energy for 3.5 TeV black hole decay events for all events (upper), for events with a remnant (middle) and for events without remnant (lower)}
\label{Et35}
\end{figure}

\begin{figure}
\center
\includegraphics[width=0.6\textwidth]
{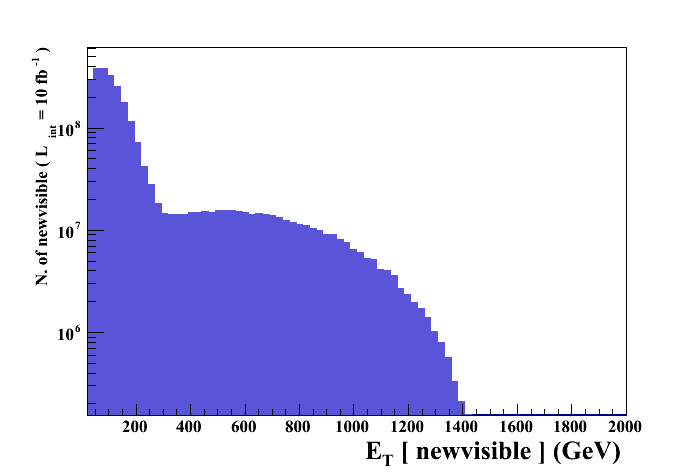}
\includegraphics[width=0.6\textwidth]
{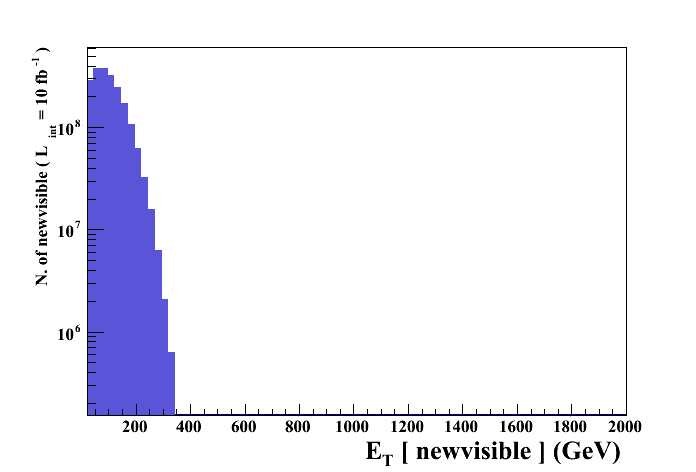}
\includegraphics[width=0.6\textwidth]
{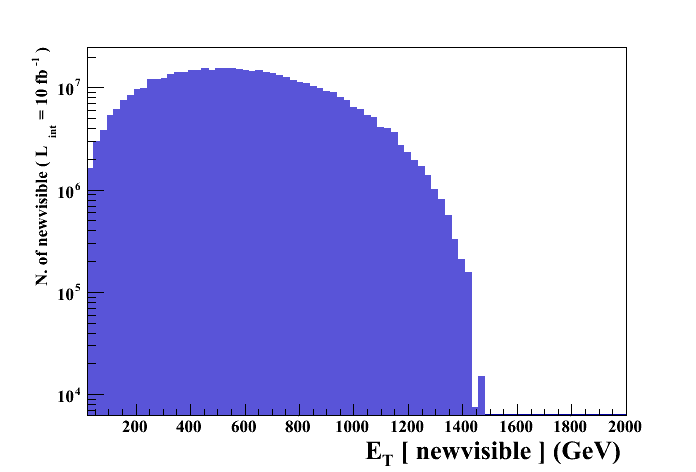}
\caption{Missing transverse energy for 3 TeV black hole decay events for all events (upper), for events with a remnant (middle) and for events without remnant (lower)}
\label{Et30}
\end{figure}

\begin{figure}
\center
\includegraphics[width=0.6\textwidth]
{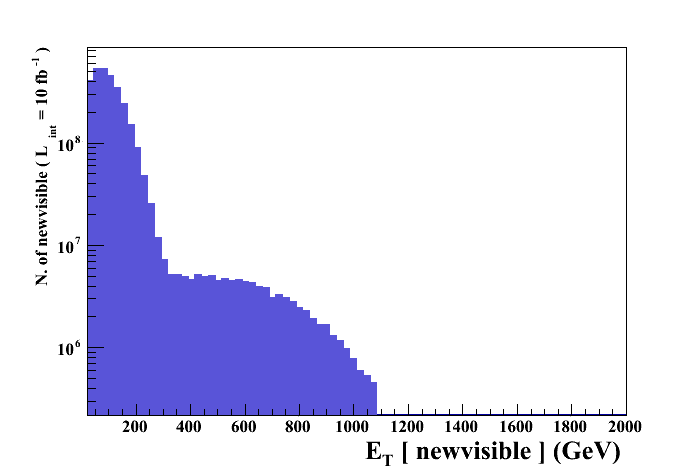}
\includegraphics[width=0.6\textwidth]
{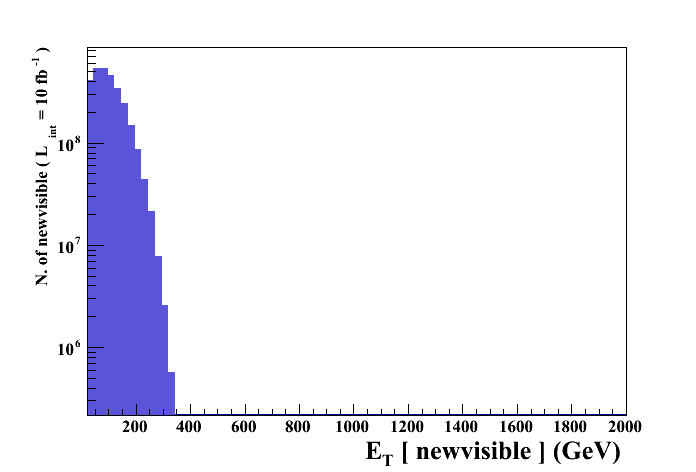}
\includegraphics[width=0.6\textwidth]
{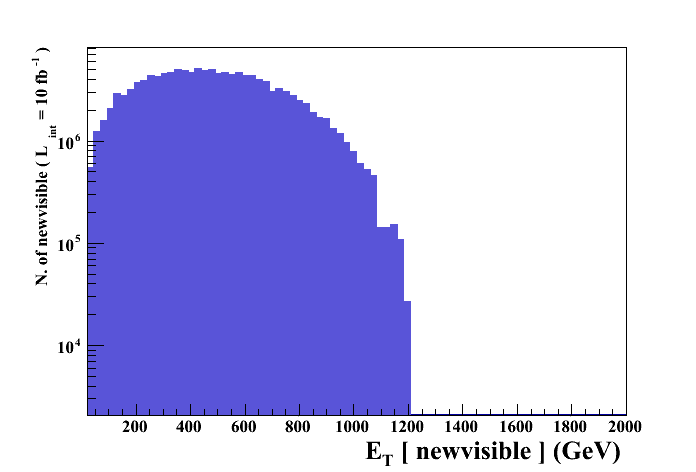}
\caption{Missing transverse energy for 2.5 TeV black hole decay events for all events (upper), for events with a remnant (middle) and for events without remnant (lower)}
\label{Et25}
\end{figure}

\begin{figure}
\center
\includegraphics[width=0.6\textwidth]
{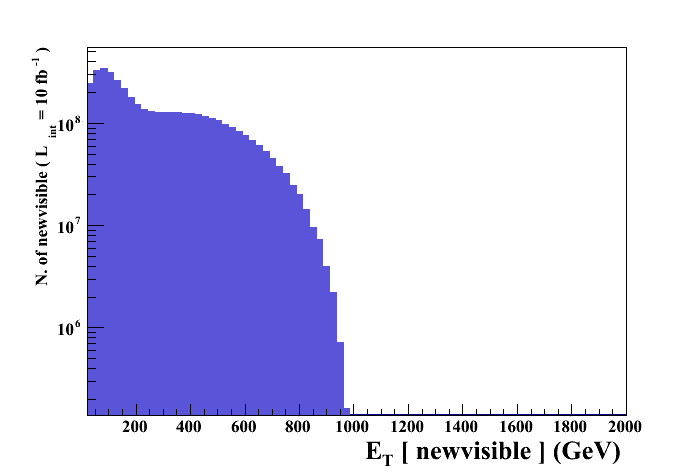}
\includegraphics[width=0.6\textwidth]
{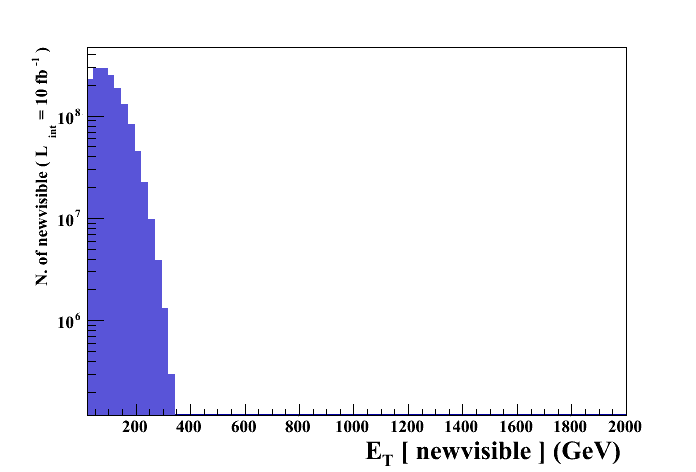}
\includegraphics[width=0.6\textwidth]
{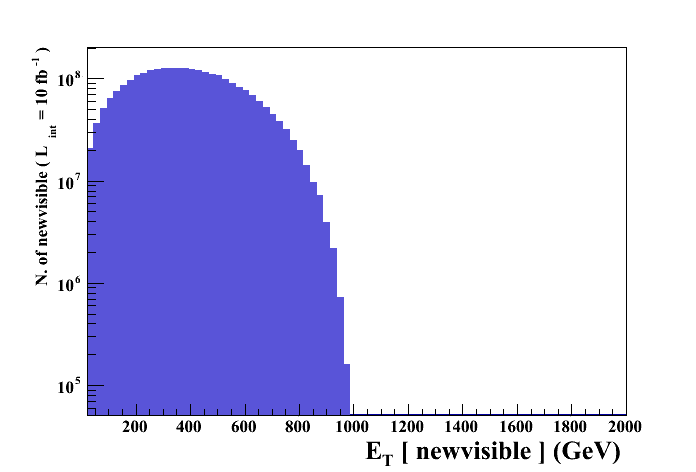}
\caption{Missing transverse energy for 2 TeV black hole decay events for all events (upper), for events with a remnant (middle) and for events without remnant (lower)}
\label{Et20}
\end{figure}

\begin{figure}
\center
\includegraphics[width=0.6\textwidth]
{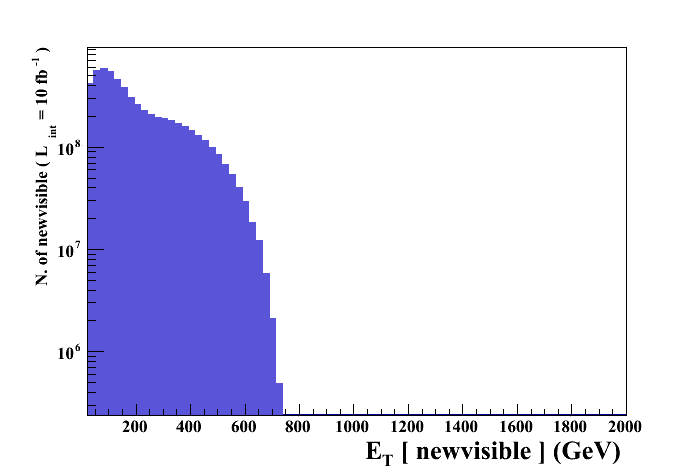}
\includegraphics[width=0.6\textwidth]
{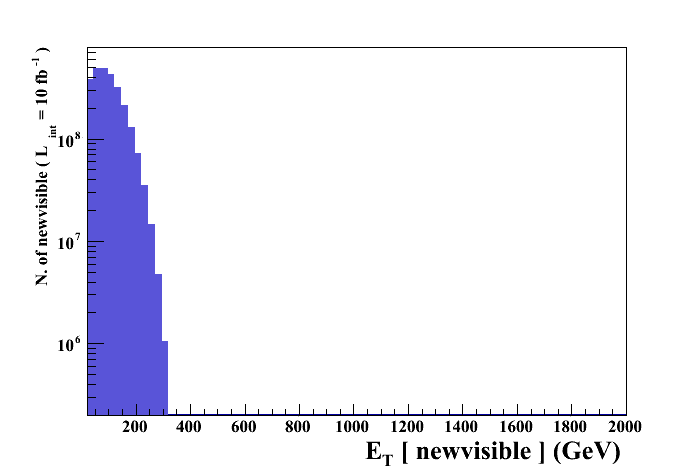}
\includegraphics[width=0.6\textwidth]
{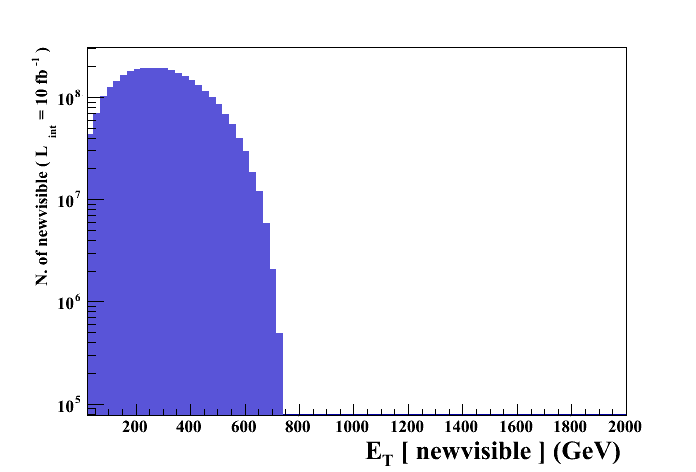}
\caption{Missing transverse energy for 1.5 TeV black hole decay events for all events (upper), for events with a remnant (middle) and for events without remnant (lower)}
\label{Et15}
\end{figure}

\begin{figure}
\center
\includegraphics[width=0.6\textwidth]
{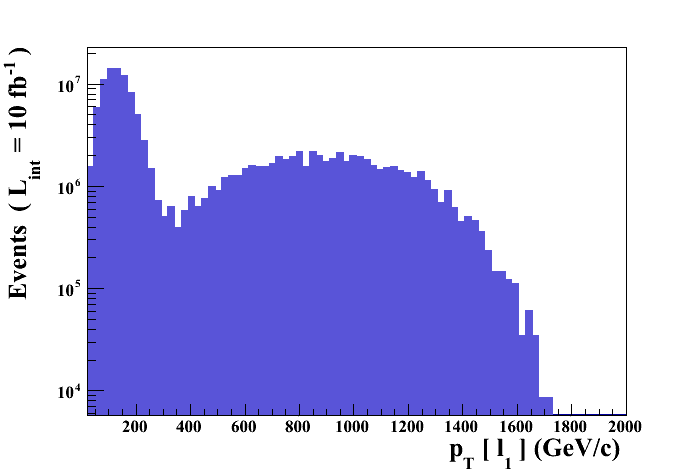}
\includegraphics[width=0.6\textwidth]
{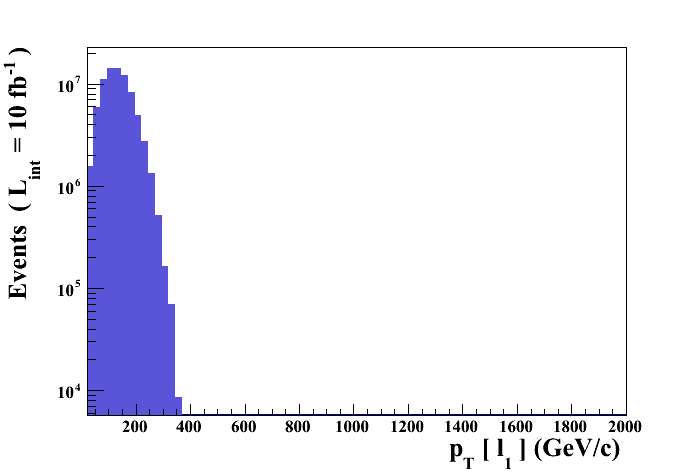}
\includegraphics[width=0.6\textwidth]
{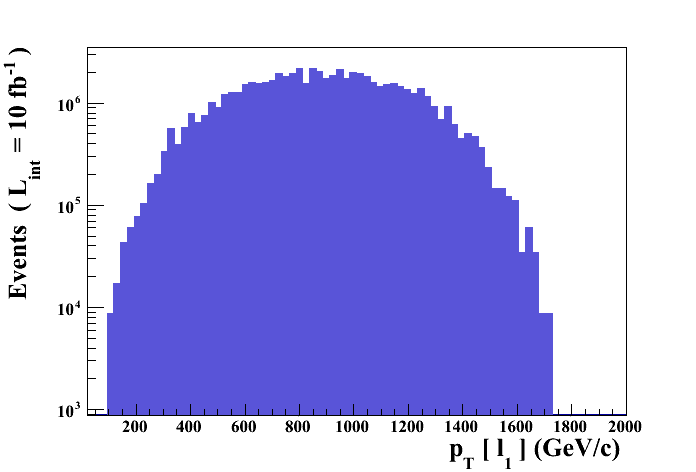}
\caption{PT(leading lepton) for 3.5 TeV black hole decay events for all events (upper), for events with a remnant (middle) and for events without remnant (lower)}
\label{PtLep35}
\end{figure}

\begin{figure}
\center
\includegraphics[width=0.6\textwidth]
{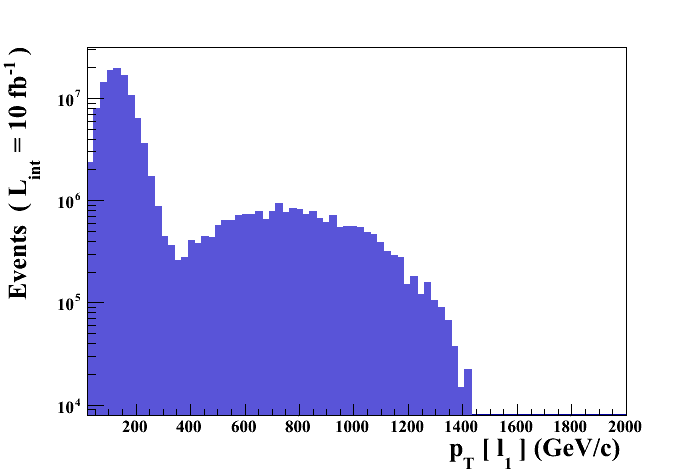}
\includegraphics[width=0.6\textwidth]
{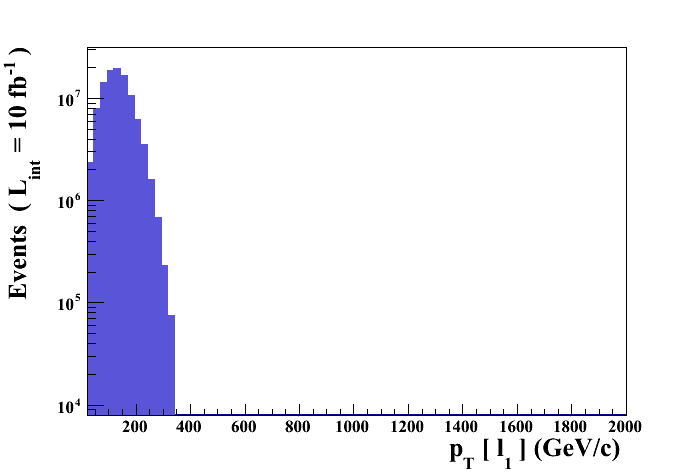}
\includegraphics[width=0.6\textwidth]
{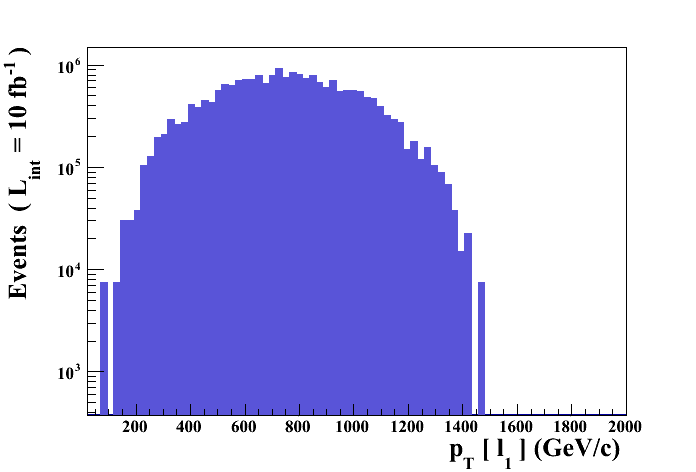}
\caption{Leading lepton transverse momentum for 3 TeV black hole decay events for all events (upper), for events with a remnant (middle) and for events without remnant (lower)}
\label{PtLep30}
\end{figure}

\begin{figure}
\center
\includegraphics[width=0.6\textwidth]
{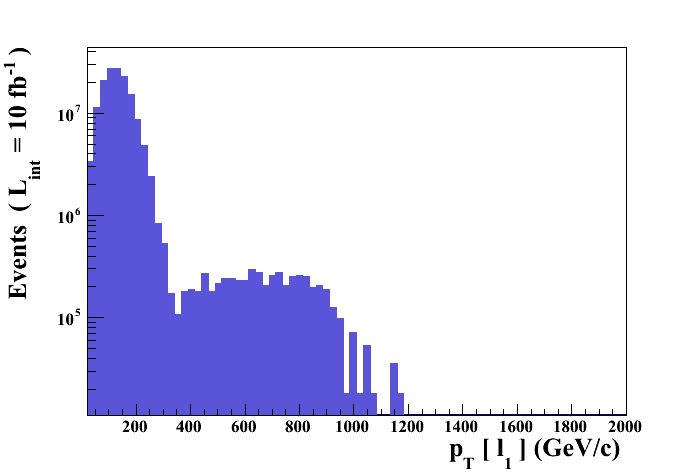}
\includegraphics[width=0.6\textwidth]
{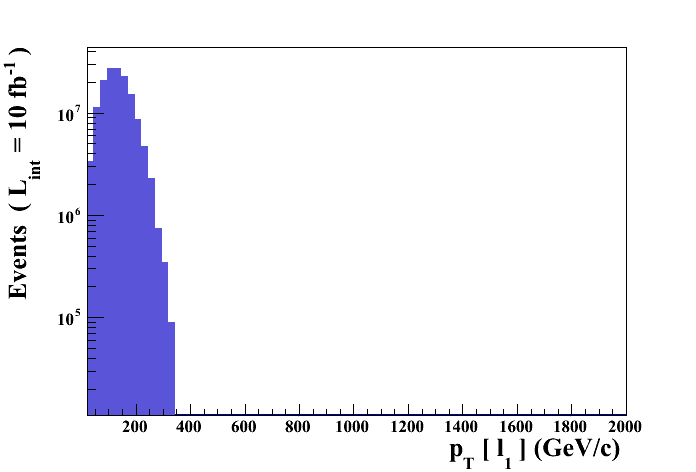}
\includegraphics[width=0.6\textwidth]
{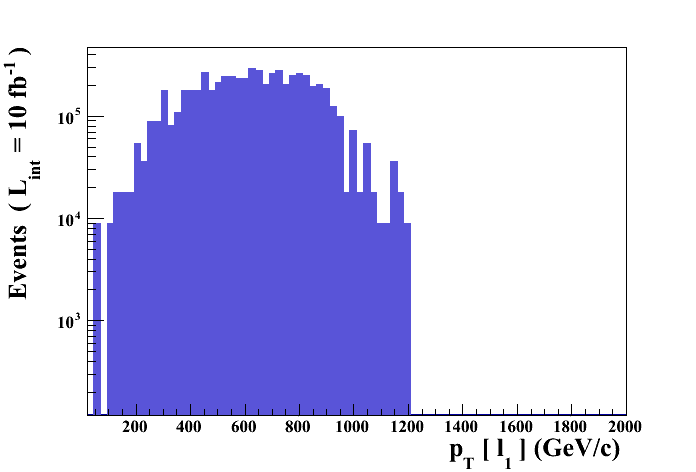}
\caption{Leading lepton transverse momentum for 2.5 TeV black hole decay events for all events (upper), for events with a remnant (middle) and for events without remnant (lower)}
\label{PtLep25}
\end{figure}

\begin{figure}
\center
\includegraphics[width=0.6\textwidth]
{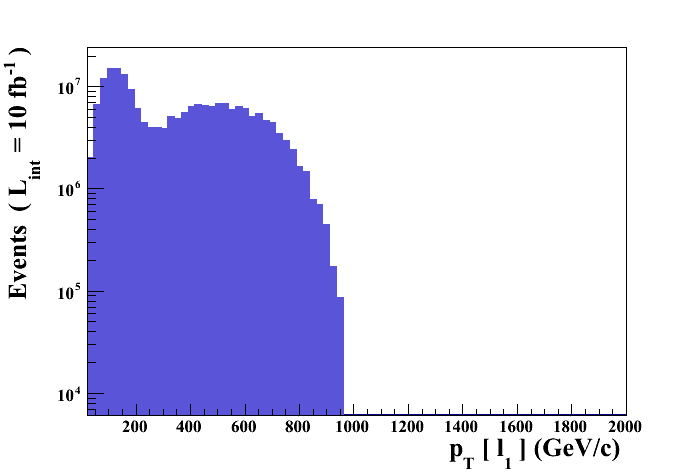}
\includegraphics[width=0.6\textwidth]
{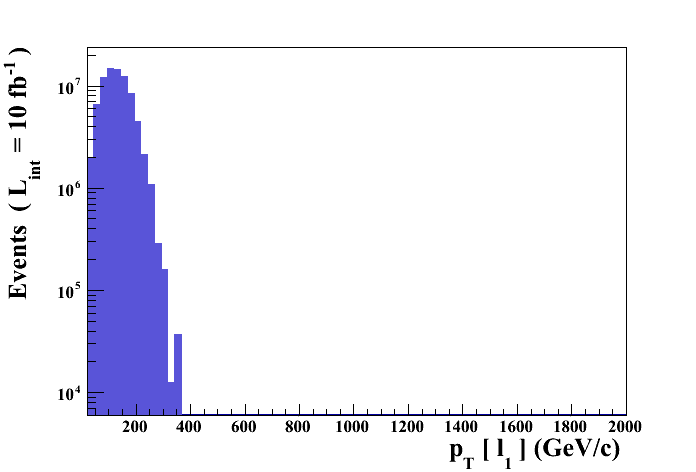}
\includegraphics[width=0.6\textwidth]
{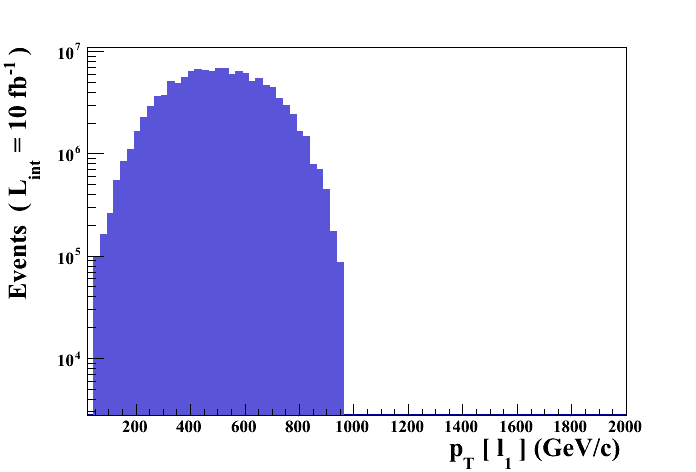}
\caption{Leading lepton transverse momentum for 2 TeV black hole decay events for all events (upper), for events with a remnant (middle) and for events without remnant (lower)}
\label{PtLep20}
\end{figure}

\begin{figure}
\center
\includegraphics[width=0.6\textwidth]
{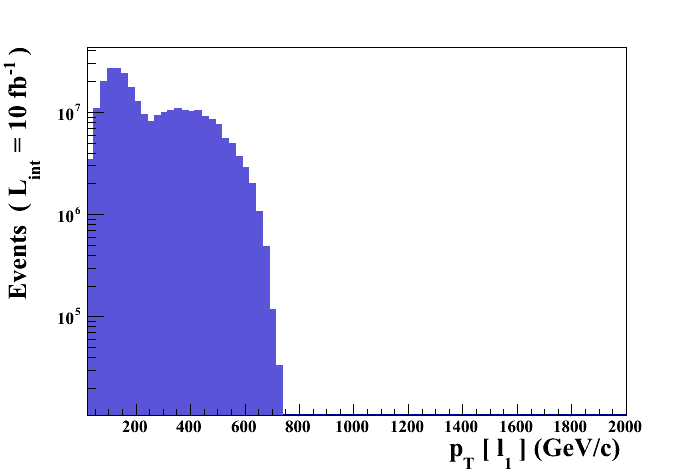}
\includegraphics[width=0.6\textwidth]
{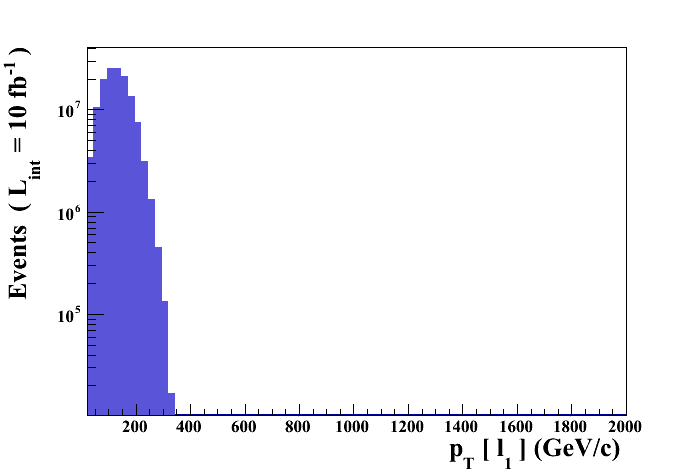}
\includegraphics[width=0.6\textwidth]
{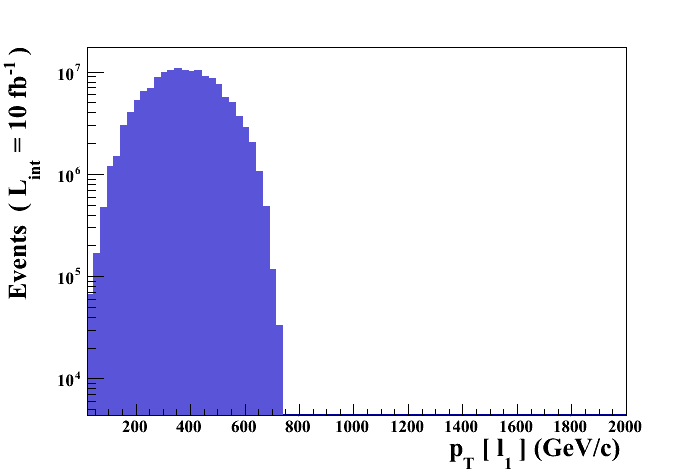}
\caption{Leading lepton transverse momentum for 1.5 TeV black hole decay events for all events (upper), for events with a remnant (middle) and for events without remnant (lower)}
\label{PtLep15}
\end{figure}

\begin{figure}[htp]
\center
\includegraphics[width=0.6\textwidth]
{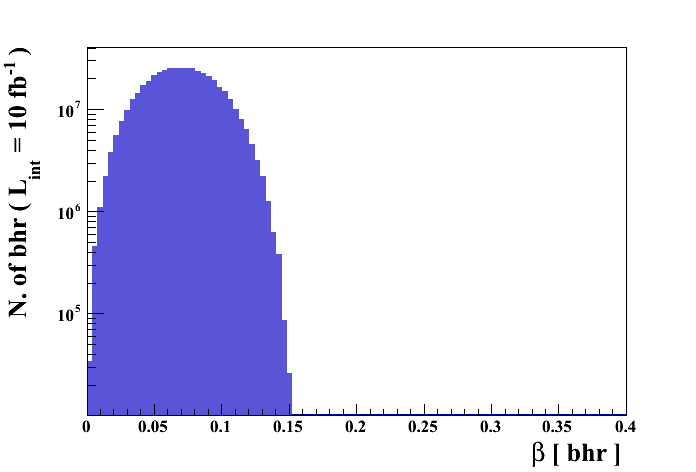}
\includegraphics[width=0.6\textwidth]
{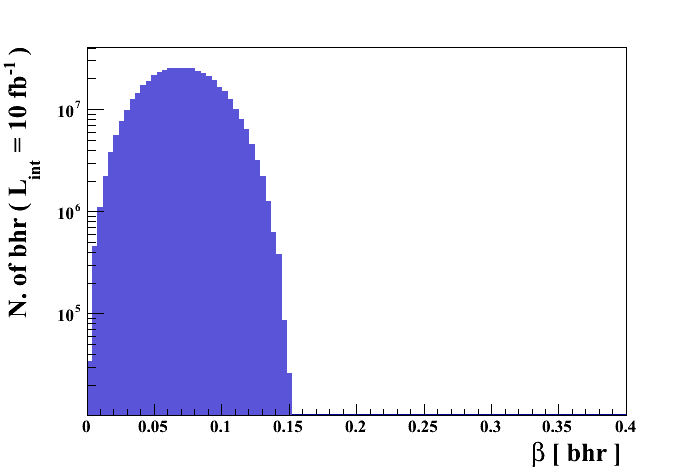}
\caption{$\beta$ of the remnantfor 3.5 TeV black hole decay events for all events (upper) and for events with a remnant (lower)}
\label{Betabhr35}
\end{figure}

\begin{figure}[htp]
\center
\includegraphics[width=0.6\textwidth]
{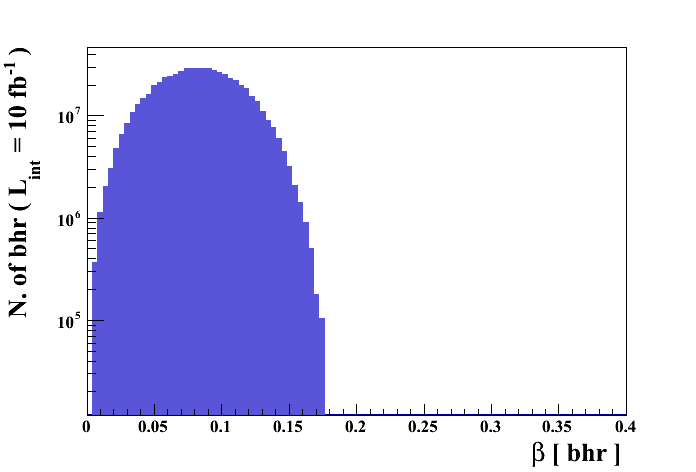}
\includegraphics[width=0.6\textwidth]
{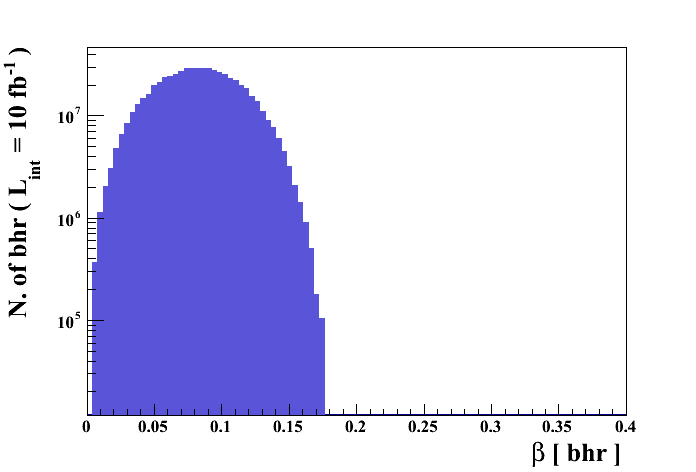}
\caption{$\beta$ of the remnant for 3 TeV black hole decay events for all events (upper) and for events with a remnant (lower)}
\label{Betabhr30}
\end{figure}

\begin{figure}
\center
\includegraphics[width=0.6\textwidth]
{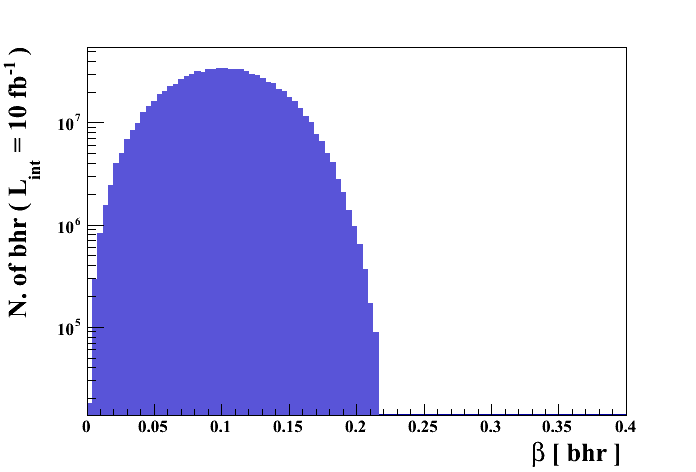}
\includegraphics[width=0.6\textwidth]
{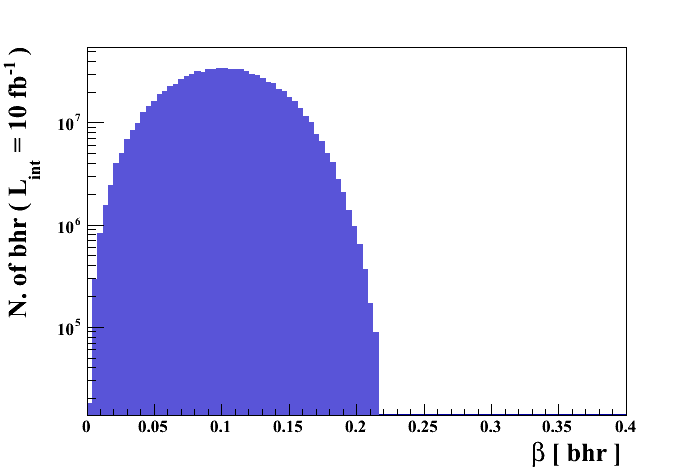}
\caption{$\beta$ of the remnant for 2.5 TeV black hole decay events for all events (upper) and for events with a remnant (lower)}
\label{Betabhr25}
\end{figure}

\begin{figure}
\center
\includegraphics[width=0.6\textwidth]
{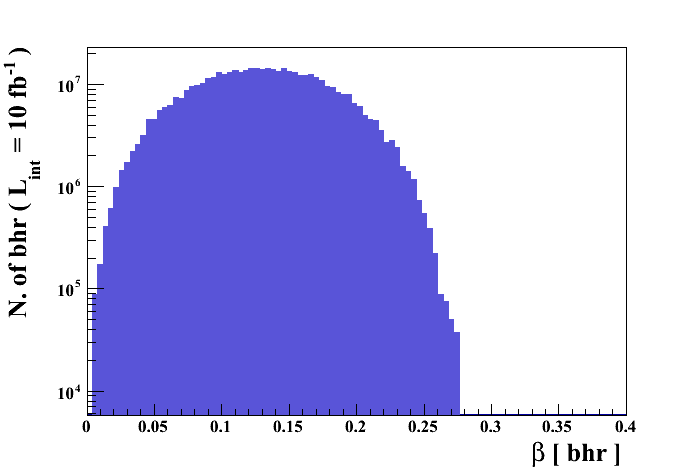}
\includegraphics[width=0.6\textwidth]
{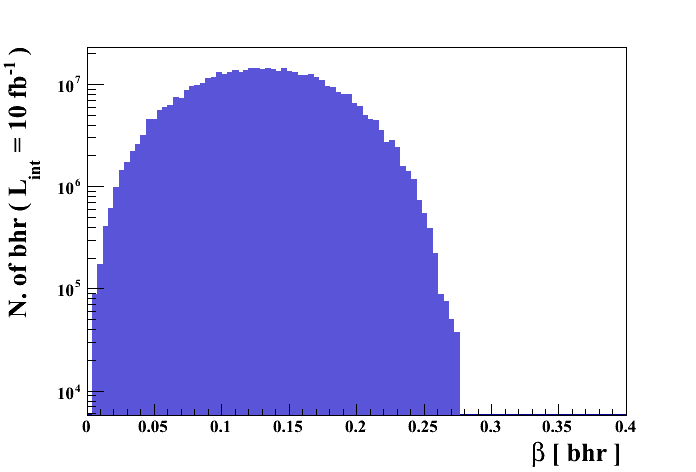}
\caption{$\beta$ of the remnant for 2 TeV black hole decay events for all events (upper) and for events with a remnant (lower)}
\label{Betabhr20}
\end{figure}

\begin{figure}
\center
\includegraphics[width=0.6\textwidth]
{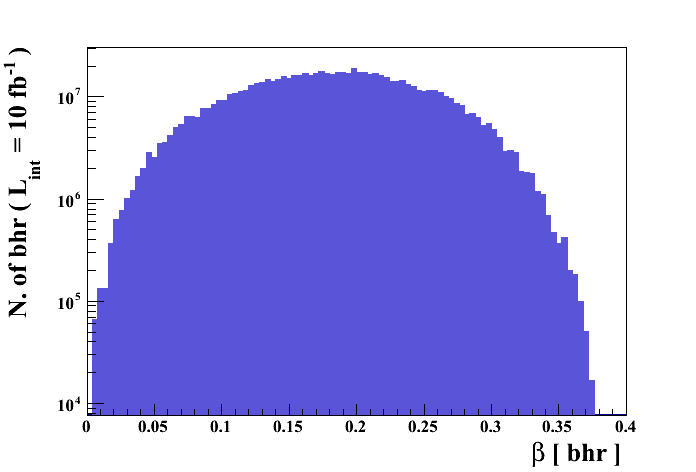}
\includegraphics[width=0.6\textwidth]
{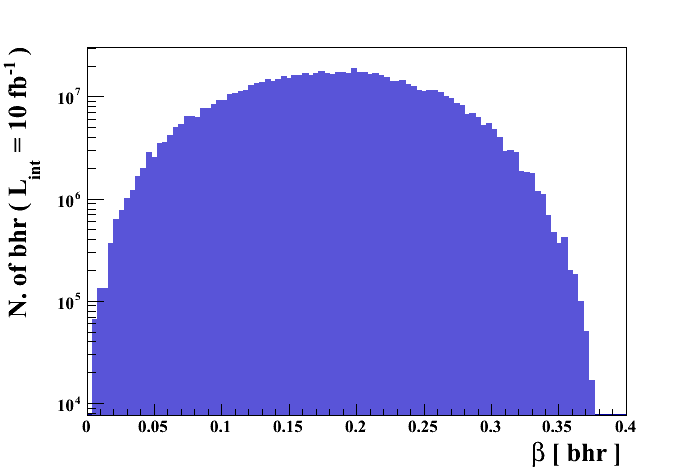}
\caption{$\beta$ of the remnant for 1.5 TeV black hole decay events for all events (upper) and for events with a remnant (lower)}
\label{Betabhr15}
\end{figure}

\setcounter{equation}{0} 
In order to analyze the decay events we employed the powerful tools provided by MadAnalysis (see \cite{Conte:2012fm}).
In fact it is possible to import the decay files produced by MadGraph and perform cuts and selections
on global and individual kinematical properties of the events.
In  particular, in order to point out the main characteristics of the different scenarios
under study, and the possibility to observe a signal over the SM background, 
we have applied a basic selection to the hadron level Monte Carlo samples using
the following event variables:
\begin{itemize}
\item
the missing transverse momentum,
\be
P_T^{\rm miss}
=
\sqrt{\left(\sum_{i} P_{x_i}\right)^2 + \left(\sum_{i} P_{y_i}\right)^2}
\ ,
\ee
where $P_{x_i}$ and $P_{y_i}$ are the cartesian $x$ and $y$ components of the
momentum of the $i^{\rm th}$ particle, and $i$ runs over all the undetectable final
state particles: neutrinos, gravitons and neutral BH remnant;
\item
the visible transverse energy,
\be 
E_T^{\rm vis}
=\sum_{k}\left(\sqrt{P_{x_k}^2+P_{y_k}^2}\right)
\ ,
\ee
where $k$ runs over all the detectable final state particles.
\item
$P_T^{\rm lep}$, defined as the transverse momentum of the leading lepton
($e$ or $\mu$) with $|\eta^{\rm lep}| < 2.5$, where $\eta^{\rm lep}$ is the lepton
pseudorapidity;
\item
$\beta(bhr)$, the velocity of the black hole remnant
\end{itemize} 
The corresponding histograms are shown in figs. (\ref{Pt35} to \ref{Betabhr15}).
These kinematical variables should be interesting as the
observation of final state particles with high transverse momentum 
is the typical signal for the decay of a large mass state.
In particular, the requirement of high-energy leptons and/or high missing
transverse momentum allows to cope with the huge QCD background at the LHC.
Since BHs decay democratically to all SM particles, the search for extremely
energetic leptons is, in this context, one of the most direct way to look for deviations 
from the SM predictions. 
\par
However, in case of stable neutral remnants, the lepton signal can be depressed
by the fact that a relevant fraction of the BH mass is not available in the decay.
On the other hand, the neutral remnant, behaving as a WIMP, will carry away a lot
of energy and enhance the missing transverse energy signal.
\par
Finally, we will see that the massive black hole remnants have a velocity much lower than that
of the standard model particles produced.
This feature was already noted in \cite{Alberghi:2013hca}, where it was also shown that the expected velocity
distributions for neutral and charged remnants looks alike (therefore providing a mechanism for the
detection and identification for charged remnants).
We plan to allow for the production of charged remnants in the following developments of the research.
Let us now briefly comment on the behavior shown in the figures.
One can first note that the high transverse momentum component due to the black hole remnant
decreases as the black hole and remnant mass gets lower (see figs. \cite{Pt35} to \cite{Pt15})
In a further development of this analysis one would want to disentangle the effect of
the black hole and remnant masses, as in the present model they are strictly related.
The same qualitative behavior appears for the visible transverse energy and for the transverse momentum
of the leading lepton (figs. \cite{Et35} to {PtLep15}), even if the amount of energy available to standard model particles 
in events containing a remnant in the final state is always 500 GeV.
This aspect deserves further investigation in the following stages of the model development.
Finally the mean velocity of the remnant gets higher as its mass gets lower, as one might have naively expected.
\par
We want to conclude by stressing a peculiarity of the results.
If one compare the results with standard black hole decay analysis, one can note that the missing transverse
momentum carried away by the neutral remnant is not as high as one could usually 
expect (see for example \cite{Koch:2005ks,Hossenfelder:2005bd}), meaning that some  
black hole decays could be hidden by background events.   

\section{Conclusions and possible developments}
\label{conclusions}
\setcounter{equation}{0}
We have begun the investigation of the decay of quantum black holes with masses of the order of the TeV,
possibly produced at the Large Hadron Collider, within the framework of MadGraph 5,
a standard particle physics Monte Carlo generator.
We have written an effective Lagrangian to describe the black hole decay, examining both the possibility of a complete decay 
in standard model particles and the production of a very massive stable neutral remnant.
The analysis was performed using the MadAnalysis tool, allowing us to point out some peculiar and somehow unexpected
features of the decay process, which need to be further inevestigated in the following stages of this research.
This will be done by allowing more flexibility in the model and considering a wider range for the parameters of the 
model, considering even the case where a charged remnant is produced.

%The existence of semi-classical remnant black holes have been the subject of  
%Monte Carlo simulations~\cite{Bellagamba:2012wz} employing the code
%CHARYBDIS2~\cite{charybdis2}.
%Overall, these remnants will have a typical speed $\beta_0=v_0/c$ with the distribution shown
%in the left panel of Fig.~\ref{beta}, where two different scenarios for the end-point
%of the evaporation were assumed.

%%% Modello di figure a coppie

%\begin{figure}[t]
%\centerline{\begin{tabular}{ r}
%%\centerline{\hbox{
%\epsfig{figure=betaremn-fullfalse-14tev_3.5_6.eps,height=6cm,clip=}
%\epsfig{figure=remnmass-fullfalse-14Tev_3.5_6.eps,height=6cm,clip=}
%}}
%\end{tabular}}
%\caption{Distribution of $\beta_0$ (left panel) and mass $M_0$ (in GeV; right panel)
%of the remnant black holes for KINCUT=TRUE (dashed line) and KINCUT=FALSE (solid line).
%Both plots are for $\sqrt{s}=14\,$TeV with $M_{\rm G}=3.5\,$TeV in $D=6$
%total dimensions and $10^4$ total black hole events.
%\label{beta}}
%\end{figure}

%
%
%\section*{Acknowledgements}
%
%

%

%
%

%

%%%%%%%%%%%%%%%%%%%%%%%%%%%%  PT for 3.5 TeV BH  %%%%%%%%%%%%%%%%%%%%%%%%%%%%

%%%%%%%%%%%%%%%%%%%%%%%%%%%%  PT for 3.0 TeV BH  %%%%%%%%%%%%%%%%%%%%%%%%%%%%

%%%%%%%%%%%%%%%%%%%%%%%%%%%%  PT for 2.5 TeV BH  %%%%%%%%%%%%%%%%%%%%%%%%%%%%

%%%%%%%%%%%%%%%%%%%%%%%%%%%%  PT for 2.0 TeV BH  %%%%%%%%%%%%%%%%%%%%%%%%%%%%

%%%%%%%%%%%%%%%%%%%%%%%%%%%%  PT for 1.5 TeV BH  %%%%%%%%%%%%%%%%%%%%%%%%%%%%

%%%%%%%%%%%%%%%%%%%%%%%%%%%%%%%%%%%%%%%%%%%%%%%%%%%%%%%%%%%%%%%%%%%%%%%%%

%%%%%%%%%%%%%%%%%%%%%%%%%%%%  ET for 3.5 TeV BH  %%%%%%%%%%%%%%%%%%%%%%%%%%%%

%%%%%%%%%%%%%%%%%%%%%%%%%%%%  ET for 3.0 TeV BH  %%%%%%%%%%%%%%%%%%%%%%%%%%%%

%%%%%%%%%%%%%%%%%%%%%%%%%%%%  ET for 2.5 TeV BH  %%%%%%%%%%%%%%%%%%%%%%%%%%%%

%%%%%%%%%%%%%%%%%%%%%%%%%%%%  ET for 2.0 TeV BH  %%%%%%%%%%%%%%%%%%%%%%%%%%%%

%%%%%%%%%%%%%%%%%%%%%%%%%%%%  ET for 1.5 TeV BH  %%%%%%%%%%%%%%%%%%%%%%%%%%%%

%%%%%%%%%%%%%%%%%%%%%%%%%%%%%%%%%%%%%%%%%%%%%%%%%%%%%%%%%%%%%%%%%%%%%%%%%%

%%%%%%%%%%%%%%%%%%%%%%%%%%%%  PT(leading lepton) for 3.5 TeV BH  %%%%%%%%%%%%%%%%%%%%%%%%%%%%

%%%%%%%%%%%%%%%%%%%%%%%%%%%%  PT(leading lepton) for 3.0 TeV BH  %%%%%%%%%%%%%%%%%%%%%%%%%%%%

%%%%%%%%%%%%%%%%%%%%%%%%%%%%  PT(leading lepton) for 2.5 TeV BH  %%%%%%%%%%%%%%%%%%%%%%%%%%%%

%%%%%%%%%%%%%%%%%%%%%%%%%%%%  PT(leading lepton) for 2.0 TeV BH  %%%%%%%%%%%%%%%%%%%%%%%%%%%%

%%%%%%%%%%%%%%%%%%%%%%%%%%%%  PT(leading lepton) for 1.5 TeV BH  %%%%%%%%%%%%%%%%%%%%%%%%%%%%

%%%%%%%%%%%%%%%%%%%%%%%%%%%%%%%%%%%%%%%%%%%%%%%%%%%%%%%%%%%%%%%%%%%%%%%%%%

\clearpage

%%%%%%%%%%%%%%%%%%%%%%%%%%%%  $\beta(bhr)$  for 3.5 TeV BH  %%%%%%%%%%%%%%%%%%%%%%%%%%%%

%%%%%%%%%%%%%%%%%%%%%%%%%%%%  $\beta(bhr)$ for 3.0 TeV BH  %%%%%%%%%%%%%%%%%%%%%%%%%%%%

%%%%%%%%%%%%%%%%%%%%%%%%%%%%  $\beta(bhr)$ for 2.5 TeV BH  %%%%%%%%%%%%%%%%%%%%%%%%%%%%

%%%%%%%%%%%%%%%%%%%%%%%%%%%%  $\beta(bhr)$ for 2.0 TeV BH  %%%%%%%%%%%%%%%%%%%%%%%%%%%%

%%%%%%%%%%%%%%%%%%%%%%%%%%%%  $\beta(bhr)$ for 1.5 TeV BH  %%%%%%%%%%%%%%%%%%%%%%%%%%%%

%%%%%%%%%%%%%%%%%%%%%%%%%%%%%%%%%%%%%%%%%%%%%%%%%%%%%%%%%%%%%%%%%%%%%%%%%%

%%%%%%%%%%%%%%%%%%%%%%%%%%%%%%%%%%%%%%%%%%%%%%%%%%%%%%%%%%%%%%%%%%%%%%%%%%

\end{document}